\documentclass[reprint,superscriptaddress,preprintnumbers,amsmath,amssymb,floatfix]{revtex4-1}
\usepackage{graphicx}
\usepackage{amsmath}
\usepackage{gensymb}
\usepackage{mathtools}
\usepackage{comment}

\begin{document}
\title{Limit cycles and bifurcations in a nonlinear MEMS oscillator with a 1:3 internal resonance. Part I: The case of a driven resonator}
\author{S. Houri} \email{Houri.Samer@lab.ntt.co.jp, samer.houri.dg@hco.ntt.co.jp}
\affiliation{NTT Basic Research Laboratories, NTT Corporation, 3-1 Morinosato-Wakamiya, Atsugi-shi, Kanagawa 243-0198, Japan}
\author{D. Hatanaka}
\affiliation{NTT Basic Research Laboratories, NTT Corporation, 3-1 Morinosato-Wakamiya, Atsugi-shi, Kanagawa 243-0198, Japan}
\author{M. Asano}
\affiliation{NTT Basic Research Laboratories, NTT Corporation, 3-1 Morinosato-Wakamiya, Atsugi-shi, Kanagawa 243-0198, Japan}
\author{R. Ohta}
\affiliation{NTT Basic Research Laboratories, NTT Corporation, 3-1 Morinosato-Wakamiya, Atsugi-shi, Kanagawa 243-0198, Japan}
\author{H. Yamaguchi}\email{ hiroshi.yamaguchi.zc@hco.ntt.co.jp}
\affiliation{NTT Basic Research Laboratories, NTT Corporation, 3-1 Morinosato-Wakamiya, Atsugi-shi, Kanagawa 243-0198, Japan}

\date{\today}

\begin{abstract}
This work investigates the behavior of an AlGaAs/GaAs piezoelectric nonlinear MEMS oscillator exhibiting a 1:3 internal resonance. The device is explored in an open-loop configuration, i.e. as a driven resonator, where depending on the drive conditions we observe energy transfer between the first and third modes, and the emergence of supercritical Hopf limit cycles. We examine the dependence of these bifurcations on the oscillator\textsc{\char13}s frequency and amplitude, and reproduce the observed behavior using a system of nonlinearly coupled equations which show interesting scaling behavior.

\end{abstract}

\maketitle

\indent\indent\ Micromechanical systems (MEMS) now pervade our lives as myriad sensors and actuators in electronic systems and equipment. In addition to their ability to react to an applied force, micro- and nano-electromechanical systems (M/NEMS) distinguish themselves as an excellent and convenient test bed for the investigation of nonlinear dynamics \cite{lifshitz2008nonlinear,rhoads2008nonlinear}.
M/NEMS have been used as platforms to observe a number of nonlinear phenomena that include Duffing response, mode coupling \cite{westra2010nonlinear,matheny2013nonlinear}, parametric amplification and oscillations \cite{rugar1991mechanical,mahboob2008bit}, stochastic resonance \cite{venstra2013stochastic}, phonon lasing \cite{mahboob2013phonon}, and oscillator synchronization \cite{matheny2014phase,antonio2015nonlinearity,houri2017direct} to name a few examples.\\
\indent\indent\ Furthermore, M/NEMS devices are particularly suited to investigate the rich dynamics of internal resonance since they demonstrate multiple nonlinear modes. Internal resonance is achieved when two resonant modes have their frequency ratio equal to the order of the nonlinearity of the system, whereby the lower frequency vibrations are up-converted to a frequency that drives the higher mode. Thus due to internal resonance energy is transferred between the two modes and the nonlinear coupling changes significantly the behavior of the system.\\
\indent\indent\ Internal resonance mediated through cubic nonlinearity has been observed in a variety of electromechanical systems including quartz crystals \cite{kirkendall2013internal} and MEMS oscillators \cite{arroyo2016duffing,antonio2012frequency,taheri2017mutual}. This behavior can be of practical value, since the finger print of internal resonance, usually seen as a plateau feature in the nonlinear response of the lower frequency mode, was suggested as a means to fix the oscillator\textsc{\char13}s frequency and improve its stability \cite{antonio2012frequency}.\\
\indent\indent\ Nevertheless, the introduction of internal resonance is not purely beneficial, since as the system becomes more complex it also becomes less predictable as demonstrated by the exotic dynamics of internal resonance in vibrating beams \cite{mangussi2016internal} or by the observation of anomalous energy decay in nanomechanical systems \cite{guttinger2017energy,shoshani2017anomalous}. In addition, instabilities in internally resonant oscillators are known to exist both in their steady state response \cite{zanette2018stability}, and as the appearance of limit cycles corresponding to supercritical Hopf bifurcations \cite{guttinger2017energy,mahboob2016hopf}.\\
\indent\indent\ Thus if internal resonance is to be leveraged in M/NEMS devices a more detailed investigation of instabilities is required, this is especially the case for Hopf bifurcations limit cycles since the study of their properties and their dependence on experimental parameters remains to be completed.\\
\indent\indent\ In this work, we focus on supercritical Hopf instabilities in a MEMS piezoelectric device exhibiting internal resonance between the nonlinear 1st and 3rd flexural modes. The device is first characterized as a driven resonator to extract the various device parameters, e.g. resonance frequencies, nonlinear and mode coupling parameters, and modal damping. The dynamics of internal resonance are thereafter explored as a function of amplitude and detuning of the 1st mode drive. In particular we explore the onset of super-critical Hopf bifurcations, and how their appearance and properties depend on the parameter space. Subsequently, in part II of this publication the resonator is placed in a gain feedback loop and driven as a self-sustained oscillator. By investigating both the open-loop and closed-loop modes of operation, we explore the impact of exchanging the damping term from positive to negative on the stability of internally resonant devices. Note that from hereon we will refer to the open-loop configuration simply as a resonator and the closed loop self-sustained configuration as an oscillator.\\
\indent\indent\ The piezoelectric MEMS device used in this work is fabricated using an 600 nm thick $\mathrm{Al_{0.3}Ga_{0.7}As/GaAs}$ heterostructure in the form of a clamped-clamped beam with a length of 150 $\mu$m, and a width of 20 $\mu$m. The GaAs layer forms a two-dimensional electron gas at the heterostructure\textsc{\char13}s interface which serves as a back electrode that is contacted through a gold-germanium-nickel eutectic alloy, while an evaporated gold layer acts as the top electrode. The gold electrodes are formed on either side of the beam\textsc{\char13}s length with partial coverage of the structure. The two sides are electrically insulated by a an etching of the beam\textsc{\char13}s midsection, thus electrodes can be addressed separately as shown schematically in Fig.~1(a). By exciting both electrodes, odd modes are efficiently transduced while even ones are suppressed.\\
\indent\indent\ Measurements are done at room temperature using an NEOARK laser Doppler vibrometer (LDV), the sample is placed in a vacuum chamber with optical and electrical access, and a chamber pressure of $\sim$ $10^{-4}$ Pa. Frequency sweeps done using an Vector Network Analyzer (VNA, ZVL3) show that at low driving amplitudes, $\sim$ 14 mV$_{\text{PP}}$, the system exhibits two linear resonance peaks in the 0-1 MHz range, Fig.~1(b). The linear response is fitted to give resonance frequencies of $f_{1}$ = 324 kHz, $f_{3}$ = 966 kHz, and quality factors of $Q_{1}$ = 1132, $Q_{3}$ = 1360 for the 1st and 3rd flexural modes respectively. Note that $f_{1}$ is only slightly detuned from $f_{3}/3$ (about 2 kHz) which is the key to enable internal resonance.\\
\indent\indent\ As the drive amplitude is increased, a nonlinear response is observed for both modes, with the fundamental mode showing a spring softening nonlinearity and the third mode a spring hardening nonlinearity, Fig.~1(b). The nonlinear responses are fitted to obtain Duffing nonlinear coefficients of $\alpha_{1} = -8.3\times10^{23}$ (rd.s$^{-1}$.m$^{-1}$)$^{2}$, and  $\alpha_{3}= 3\times10^{24}$ (rd.s$^{-1}$.m$^{-1}$)$^{2}$ for the first and third modes respectively \cite{suppinf}. Duffing parameter fits take into account nonlinear damping as an amplitude dependent quality factor, as done in literature for graphene nanomechanical systems \cite{davidovikj2017nonlinear}, and therefore do not require the introduction of a new parameter other than the Duffing nonlinearity.
As the drive amplitude is further increased, the nonlinear response of the first mode exhibits features such as dips, plateaus and oscillations in its frequency response sweep, Fig.~1(b), the appearance of these features is equated with the onset of internal resonance \cite{kirkendall2013internal,mangussi2016internal,guttinger2017energy}. Experimentally, such features are observed when the first mode drive amplitude exceeds 0.22 V$_{\text{PP}}$ (for a forward frequency sweep).\\
\begin{figure}[t]
	\graphicspath{{Figures/}}
	\includegraphics[width=85mm]{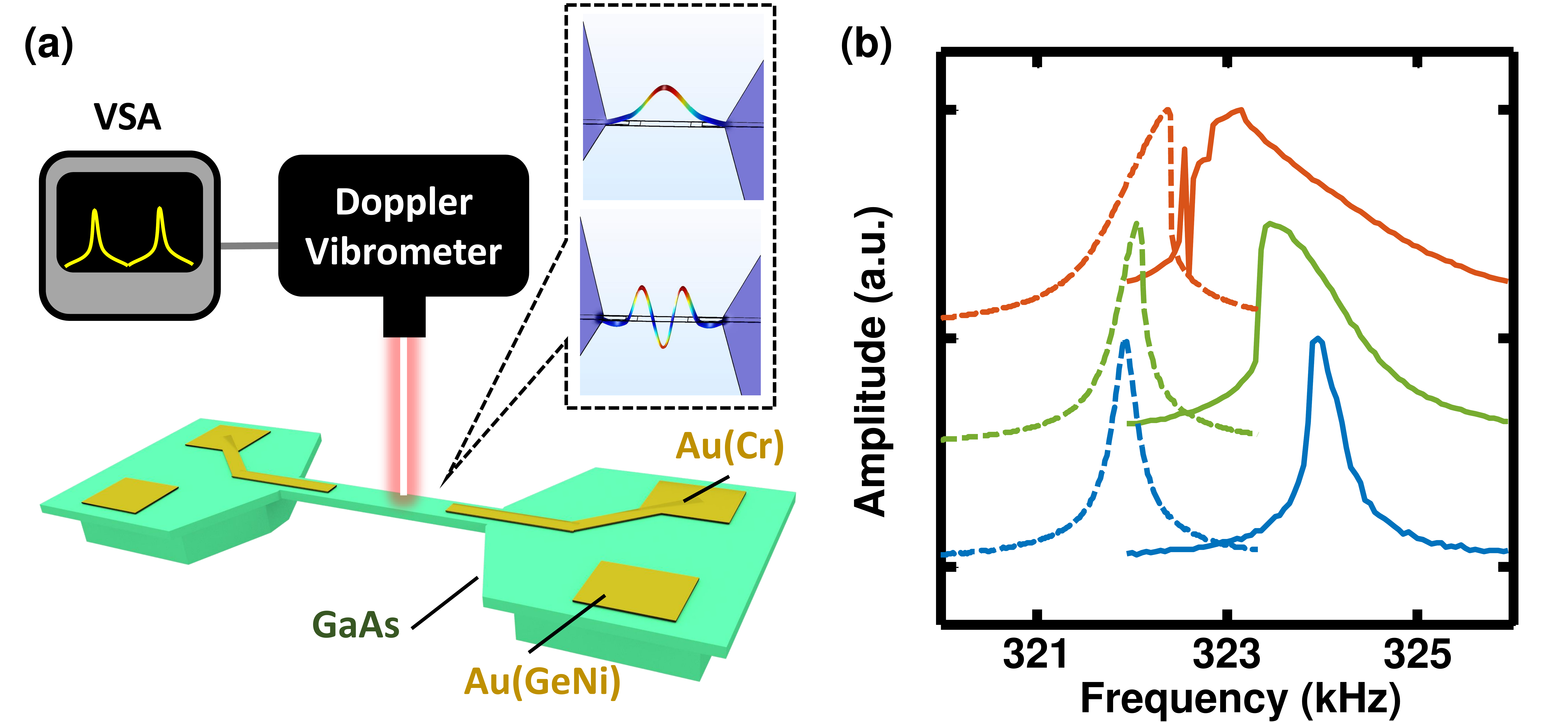}
	\caption{(a) Schematic representation of the experimental setup, a clamped-clamped beam GaAs sample is excited electrically and its vibration probed via a laser Doppler vibrometer (LDV) whose signal is sampled using an HP89410A vector signal analyzer (VSA). (b) The frequency response of the first and third modes shown as a solid and dashed lines respectively, for drive voltages of 14 mV (blue), 70 mV (green), 220 mV(red). The frequency axis for the third mode is divided by 3 to make the 1:3 ratio visible.}
\end{figure}
\indent\indent\ In addition to internal resonance we also observe dispersive mode coupling, where the vibration of one mode affects the resonance frequency of the other via added tension in the structure. The mechanics of dispersive mode coupling, result in a quadratic relation between the frequency shift ($\delta\!f$) of one mode and the amplitude of vibration of the other \cite{westra2010nonlinear,matheny2013nonlinear,lulla2012nonlinear}. We quantify the dispersive mode coupling terms $\alpha_{13}$  ($\alpha_{31}$) by strongly driving mode 3 (mode 1) with a waveform generator (WF1974), and probing mode 1 (mode 3) at low power $\sim$ -50 dBm using the VNA. The amplitude of the drive mode is equally observed using a spectrum analyzer (HP 89410A).\\
\indent\indent By sweeping the amplitude and frequency of the drive mode it is possible to reconstruct the quadratic dependence, which upon fitting a second order polynomial gives $\alpha_{13}=1.28\times10^{19}$ (rd.s$^{-1}$.m$^{-1}$)$^{2}$, as shown in Fig.~2(a). A similar value for $\alpha_{31}$ is difficult to fit accurately since internal resonance kicks-in at relatively low drive amplitudes. However, it follows from mode coupling mechanics that $\alpha_{13}$=$\alpha_{31}$ \cite{westra2010nonlinear} (once the modal effective mass is taken into account). Note that in the literature investigating internal resonance in M/NEMS the dispersive mode coupling term is sometimes included \cite{taheri2017mutual}, while it is dropped in others \cite{guttinger2017energy,shoshani2017anomalous}.\\
\indent\indent\ The modal interaction Hamiltonian, i.e. dispersive mode coupling and internal resonance, is expressed as: ${\mathcal{H}}=\frac{1}{2} \alpha_{13}'x_1^2x_3^2+gx_1^3 x_3$, where $x_1$ and $x_3$ are the modal displacements. Defining $\alpha_{13}$= $\alpha_{13}'/M_1$ and $\alpha_{31}$= $\alpha_{31}'/M_3$ where $M_1$ and $M_3$ are the modal masses, and since M$_1$ $\approx$ M$_3$, it is assumed that $\alpha_{31}$=$\alpha_{13}$. And $g$ is the internal resonance mode coupling parameter. We further define $\gamma_1$ and $\gamma_3$ as the line width of the resonant modes, and $F$ and $\omega$ as the driving force magnitude and frequency, respectively.\\
\indent\indent\ The dynamics of the system are modeled by applying the rotating frame approximation  \cite{greywall1994evading}, whereby the solution takes the form $\mathrm{x(t)=A(t)cos(\omega t+\phi(t))}$, where $\mathrm{A(t)}$ and $\mathrm{\phi(t)}$ are slowly varying amplitude and phase envelopes. By injecting the assumed solution into the governing equations, and dropping all frequency terms that are not on the order of  $\omega_1$ and $\omega_3$ for the first and third mode equations respectively, the following rotating frame dynamics are obtained:\\
\begin{widetext}
\begin{eqnarray}
\begin{rcases}
\mathrm{-2\omega_1^2\delta A_1 + i\omega_1\gamma_1 A_1 + \frac{3}{4}\alpha_1 A_1^3 + \frac{1}{2}\alpha_{13}A_3^2A_1    + \frac{3}{4}gA_1^2A_3e^{(i\Delta\phi)} + i2\omega_1\dot{A}_1 -2\omega_1A_1\dot{\phi}_1 = {\it{F}}e^{-i\phi_1}  }\\
\mathrm{ 18\omega_1^2(\Delta-\delta)A_3 + i3\omega_1\gamma_3A_3 + \frac{3}{4}\alpha_3A_3^2 + \frac{1}{2}\alpha_{13}A_1^2A_3 + \frac{1}{4}gA_1^3e^{-i\Delta\phi} + i6\omega_1\dot{A}_3 -6\omega_1A_3\dot{\phi}_3 =0}\\
\end{rcases}
\end{eqnarray}
\end{widetext}

\indent\indent\ 
where $\delta=(\omega-\omega_1 )/\omega_1$  , $\Delta = (\omega_3- 3\omega_1 )/3\omega_1$, $\Delta\phi = \phi_3 - 3\phi_1$, and all higher order terms are dropped.
Setting the time derivative terms in system of Eq.~(1) to zero gives the steady state response of the system. In particular the second line of Eq.~(1) gives a sixth order polynomial that relates the two steady states vibration amplitudes, i.e. $A_1$ and $A_3$, as follows:\\
\begin{figure}
	\graphicspath{{Figures/}}
	\includegraphics[width=85mm]{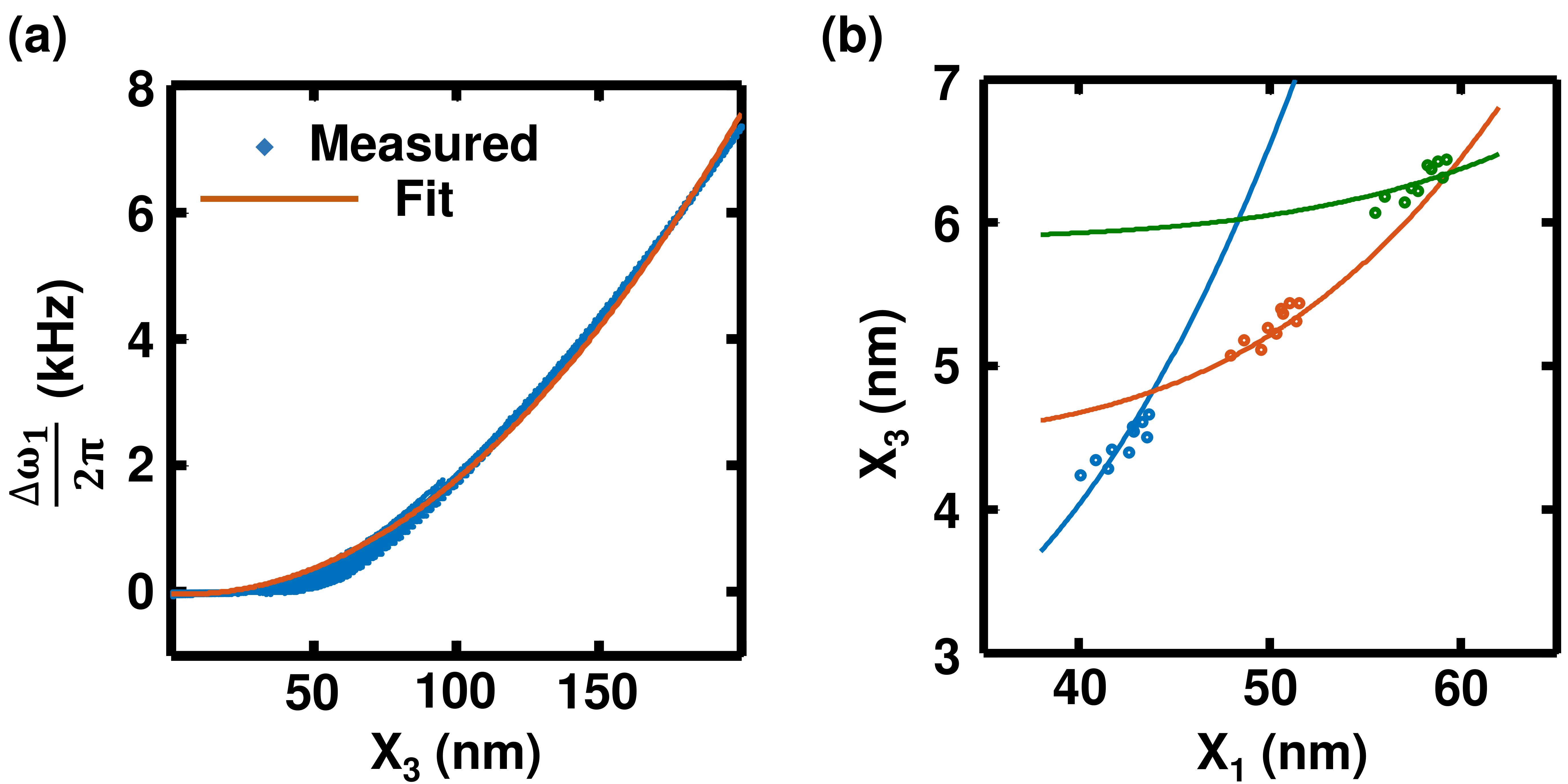}
	\caption{Fitting of the nonlinear coupling parameters as shown in (a) for the dispersive mode coupling parameter ($\alpha_{13}$). (b) Fits for the internal resonance mode coupling parameter, where only three data sets are shown for clarity corresponding to detunings of $\frac{\delta}{\Delta}$ = 3 (blue), 4.2 (red), 5.4 (green).}
\end{figure}
\begin{figure}
	\graphicspath{{Figures/}}
	\includegraphics[width=85mm]{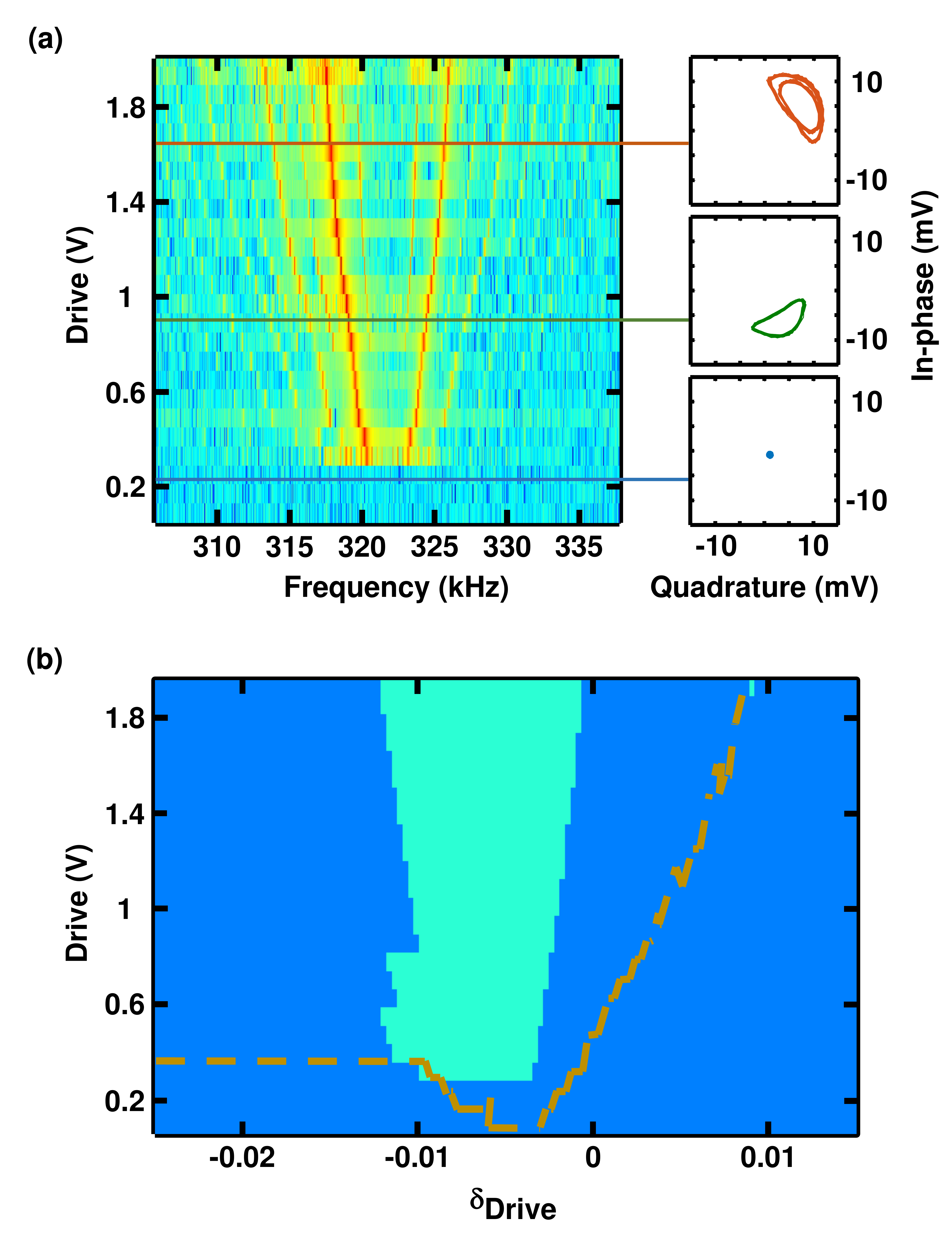}
	\caption{Frequency and phase plot response of the resonator to a 321.7 kHz drive (a). For low drive voltages ($<$ 0.33 V) the system respond as a simple nonlinear resonator (blue trace), for higher drive voltages sidebands emerge indicating the onset of a Hopf limit cycle as seen in the phase space plot (green trace), upon further increasing the drive voltage a period doubling bifurcation takes place (red trace). (b) Parameter space map, with the light areas indicating the region where a Hopf bifurcation takes place, note the existence of a feature between 0.4 and 0.8 V that breaks the otherwise symmetric shape of the bifurcation area. The dashed yellow line delineate the area where internal resonance takes place.}
\end{figure}
\begin{figure}
	\graphicspath{{Figures/}}
	\includegraphics[width=85mm]{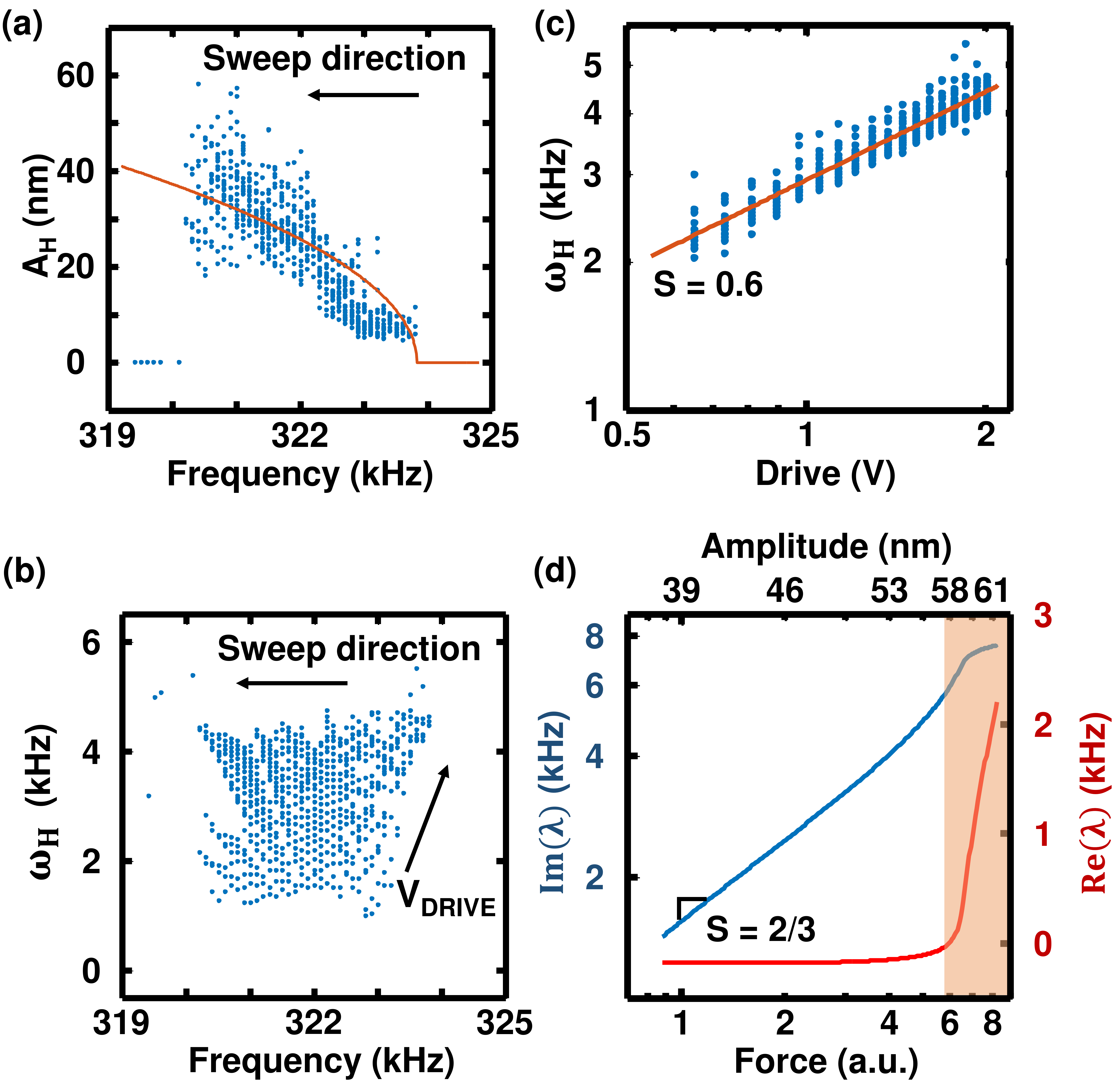}
	\caption{Scaling properties around the critical point for the (a) Hopf limit cycle amplitude as a function of the drive detuning (blue data points) with a square-root scaling power law also shown as a guide to the eye (red), the spread in experimental data prohibits a precise demonstration of the $\sim \mathcal{O}(\epsilon^{1/2})$ behavior. (b) Hopf limit cycle frequency as a function of detuning showing no dependence, hence an $\sim \mathcal{O}(1)$ scaling behavior. Also shown is the drive frequency sweep direction. (c) Hopf limit cycle frequency as a function of drive amplitude on a log-log scale demonstrating an $\sim \mathcal{O}(F^{0.6})$ response. (d) Simulation results for $\delta=0.1\Delta$ on a log-log plot showing the onset of Hopf bifurcation (shaded area), where the real component takes on positive values for a vibration amplitude $>$ 57 nm, and a slope of 2/3 for the imaginary component that changes around the Hopf bifurcation.}
\end{figure}
\begin{multline}
(18\omega_1^2(\Delta-\delta)A_3 + \frac{3}{4}\alpha_3A_3^2 +\frac{1}{2}\alpha_{13}A_1^2A_3)^2 \\
+ (3\omega_1\gamma_3A_3)^2 =   (\frac{1}{4}gA_1^3)^2 
\label{eqn:Eq2}
\end{multline}
\indent\indent\ Equation 2 indicates that energy will be transferred from the 1st to the 3rd mode without directly driving the latter. This is shown in Fig.~2(b) where the third mode amplitude is traced as a function of the first mode amplitude. Using Eq.~(2) it is possible to fit the experimental data, also shown in Fig.~2(b), and obtain the only remaining free parameter, $g = 4\times10^{26}$ (rd.s$^{-1}$.m$^{-1}$)$^{2}$.\\
\indent\indent\ Supercritical Hopf bifurcations are equally observed in the spectral domain as sidebands around the drive tone as shown in Fig.~3(a), where once the drive amplitude crosses a threshold (in this case V$_{\text{threshold}}$ = 0.33 V, for $\omega_d$ = 2$\pi\times 321.7$ kHz) sidebands emerge. Note that the frequency, i.e. spacing, of the sidebands depends on the drive amplitude. Time domain measurements of the bifurcation dynamics are undertaken using the VSA with the data shown in the in-phase versus quadrature plots in the right-side panels of Fig.~3(a). The phase-space plots show that the response of the system below the threshold is given by that of a driven resonator, and once the threshold is crossed a limit cycle emerges around the fixed point (in the rotating frame) indicative of the bifurcation. Also shown in Fig.~3(a) is the onset of period doubling bifurcation of the limit cycle itself for large drive amplitudes.\\
\indent\indent\ By sweeping the drive frequency as well as the drive amplitude it is possible to reconstruct a bifurcation diagram that shows the boundary of the area where sidebands emerge, as shown in Fig.~3(b). Note the existence of a feature in the bifurcation map located between 0.4 and 0.8 V drive amplitudes that breaks the symmetric shape of the bifurcation area. At this moment, it is not clear whether this feature is a side effect of long term measurements (i.e. drift), or an additional nonlinearity that needs to be further explored. Also note that it is possible to have energy transfer from the 1st to the 3rd mode (we define it as $A_3 > 0.1$ nm) without necessarily having Hopf bifurcations as shown in Fig.~3(b).\\
\indent\indent\ In dynamical systems it is known that different types of limit cycles scale differently around bifurcation points  \cite{strogatz2018nonlinear}, yet such scaling behavior is not often demonstrated experimentally. For instance a Hopf limit cycle is born with an  $\sim \mathcal{O}(1)$ oscillation frequency, and thereafter the frequency shows weak or no dependence on the critical parameter. The limit cycle amplitude on the other hand is given by $\sim \mathcal{O}(\epsilon^{1/2})$, i.e. the limit cycle amplitude is vanishingly small at first, but grows with a square-root scaling as the bifurcation point is crossed.\\
\indent\indent\ Here, two parameters are critical to the emergence of limit cycles, these are the drive amplitude and detuning. Thus we investigate the scaling properties of the limit cycles as a function of these parameters. Figure~4(a) shows the Hopf limit cycle amplitude ($A_H$) plotted as a function of the drive frequency (the drive frequency is swept down so as to be on the upper branch of the 1st mode response). Sweeping the drive frequency reveals that the bifurcation takes place around 324 kHz (the exact value depends on the drive amplitude), and as expected the limit cycle amplitude starts relatively small and grows as the frequency is swept beyond the bifurcation point. Although this confirms a scaling property of the form $\omega_{H}\sim \mathcal{O}(\epsilon^{a})$, where $\epsilon$ is the detuning, nevertheless the scatter in experimental data, which originates from small drift in the laser spot position, makes it difficult to fit an accurate power-law relation and thus verify the ($a = 1/2$) exponent experimentally.\\
\indent\indent\ On the other hand plotting the Hopf frequency ($\omega_H$), which is not affected by drift in laser position, as a function of detuning as shown in Fig.~4(b), reveals that upon emerging, the limit cycle has a frequency between 2 and 4 kHz (depending on the drive amplitude), but otherwise shows no significant dependence on detuning as the drive frequency is swept, thus exhibiting the $\mathcal{O}(1)$ behavior expected for Hopf bifurcations.\\
\indent\indent\ The other critical parameter for onset of bifurcation is the drive amplitude. Figure~4(c) shows a logarithmic plot of the Hopf limit cycle frequency ($\omega_H$) as a function of drive voltage. It is possible to fit a linear relation (corresponding to a power law) with an $S \approx 0.6$ exponent. The 0.6 exponent at first seems to contradict the canonical scaling properties of limit cycles since it is neither $\mathcal{O}(1)$ nor $\mathcal{O}(F^{1/2})$. This is not the case, since in textbook examples of Hopf bifurcations, a drive term is not normally included. In fact Ref. \cite{righetti2006dynamic} showed that a forcing term changes the frequency of Hopf limit cycles, and an experimentally observed power-law scaling of $S \approx 2/3$ has been reported for a forced Duffing-van der Pol oscillator \cite{houri2017direct}.\\
\indent\indent\ To further verify these results we perform numerical simulations using the rotating frame system of Eq.~(2), where the experimentally fitted parameters are injected into the model, and first the steady state solutions are calculated for a detuning of $\delta = 0.1\Delta$, and for a wide range of forcing terms. Thereafter, the Jacobian matrix is obtained for the system of equation around the steady state solutions \cite{suppinf}. It is interesting to note that the modal damping rates, given by the diagonals of the Jacobian matrix, remain unchanged for the third mode, i.e. damping rate = $-\gamma_3/2$, while the damping rate of the first mode is now modified and is given by ($-\gamma_1/2 - 9\frac{A_3^2}{A_1^2}\gamma_3$), which explains the reported anomalous energy decay in internally resonant nanomechanical systems \cite{guttinger2017energy}.\\
\indent\indent\ The real and imaginary components of the eigenvalues ($\lambda$) of the Jacobian matrix represent the damping rate and frequency response of the system respectively. The numerically calculated eigenvalues are plotted in Fig.~4(d) as a function of the forcing term and amplitude. For relatively small displacements, i.e. $<50$ nm, the simulations show that the system is damped, i.e. the real component is negative and no sidebands exist, and that the imaginary component increases with a slope of 2/3. When the steady state amplitude crosses 57 nm the real component becomes positive indicating the onset of a Hopf-bifurcation. As the system crosses this threshold the imaginary component equally changes, and the slope first steepens and then flattens. Thus simulation results demonstrate that the steady state solutions become unstable undergoing a Hopf bifurcations for large enough vibration amplitudes, while at the same time demonstrating that the Hopf frequency does scale with drive amplitude with an exponent of $\approx2/3$ as observed experimentally.\\
\bibliographystyle{apsrev}
\bibliography{aipsamp}
\end{document}